\documentclass{article}
\usepackage{graphicx}% Include figure files
\usepackage{color}
\usepackage{hyperref}
\usepackage{natbib}

\usepackage{geometry}
\hypersetup{colorlinks=false,linkcolor=blue}

\begin{document}
\bibliographystyle{aipnum4-1}
\setcitestyle{numbers,square}

\title{State space reconstruction of spatially extended systems and of time delayed systems from the time series of a scalar variable}
\author{C. Quintero-Quiroz,
M. C. Torrent,
C. Masoller.\\
Universitat Polit\`ecnica de Catalunya, Departament de F\'isica, \\Rambla St. Nebridi 22, 08222 Terrassa, Barcelona, Spain.}%

\date{\today}

% \keywords{time delay systems, high dimensional systems, space-time representation}
\maketitle

\begin{abstract}
The space-time representation of high-dimensional dynamical systems that have a well defined characteristic time scale has proven to be very useful to deepen the understanding of such systems and to uncover hidden features in their output signals.
By using the space-time representation many analogies between one-dimensional spatially extended systems (1D SESs) and time delayed systems (TDSs) have been found, including similar pattern formation and propagation of localized structures.
An open question is whether such analogies are limited to the space-time representation {or, it is also possible to recover similar evolutions in a low-dimensional pseudo-space.
To address this issue we analyze a 1D SES (a bistable reaction-diffusion system), a scalar TDS (a bistable system with delayed feedback) and a non-scalar TDS (a model of two delay-coupled lasers).}
In these three examples we show that we can reconstruct the dynamics in a three-dimensional phase space, where the evolution is governed by {the same} polynomial potential.
We also discuss the limitations of the analogy between 1D SESs and TDSs. 
\end{abstract}

\begin{quotation}
Real-world systems in physics, chemistry, biology, economy, etc. are typically described by a large number of equations, involving many variables, and therefore, their dynamical evolution occurs in a high dimensional phase space. 
One of the most exciting discoveries in the field of dynamical systems in the last decades is that, in spite of their high dimensionality, these systems can be described by low-dimensional attractors, which can be reconstructed even if one can only observe one variable, during a finite time interval, with finite resolution and with large measurement noise. 
Examples of such high dimensional systems are one-dimensional spatially extended systems (1D SESs), and time delayed systems (TDSs). 
In a space-time representation, these systems show similar phenomena (e.g., wave propagation, pattern formation, defects and dislocations, turbulence, etc.). 
In this work we study the state space reconstruction of these systems, from the time series of one scalar ``observed'' variable. {We analyze a bistable reaction-diffusion 1D SES and two TDSs: a bistable scalar system with delayed feedback, and a system composed by two lasers with delayed mutual cross coupling (the system has several variables and two time-delay terms).}
We find that their dynamics can be reconstructed in a three-dimensional pseudo space, where the evolution is governed by {the same} polynomial potential.

\end{quotation}

\section{Introduction}

The space-time representation of a high-dimensional dynamical system, by which a characteristic time-scale is used as a ``space-like dimension'', while the evolution during many characteristic times, occurs in a ``temporal dimension'', first proposed by Arecchi and co-workers in the 90s\cite{Arecchi1992,Giacomelli1996}, has proven to be extremely useful to uncover hidden space-like features, such as wave propagation, pattern formation, defects and dislocations, turbulent phenomena, etc.\cite{boccaletti1997control,maza1998control,hikihara1999expansion,nizette2003front,giacomelli2012coarsening,turitsyna2013laminar,larger2013virtual,churkin2015stochasticity}.
For example, in semiconductor lasers with time-delayed feedback, the space-time representation of the intensity time series {using the time interval between zero and the delay time} as ``space-like dimension'' has uncovered the presence of various types of
space-like structures\cite{masoller1997spatiotemporal,marino2014front,yanchuk2014pattern,marconi2015vectorial,javaloyes2015arrest,garbin2017interactions,marino2017pseudo}.

The analogy between time-delayed systems (TDSs) and one-dimensional spatially extended systems (1D SESs) is based on well-known properties, like the dimension of the attractor, which in TDSs grows linearly with the delay time, $\tau$, and the Lyapunov spectrum, which rescaled to $\tau$, is independent of $\tau$ \cite{farmer1982chaotic}.
These features correspond to the independence of the system size found in SESs. 
It is then natural to ask whether such analogies also apply to their underlying attractors. 

TDSs are infinite-dimensional systems because, in order to obtain a solution, one needs to specify as initial condition a function in the interval $[-\tau,0]$ \cite{erneux2009applied}.
However,  their dynamical evolution often occurs in low-dimensional attractors.
If the TDS is described by a scalar delay-differential equation of the form  $\dot{u}=f(u,u(t-\tau))$, in a three-dimensional pseudo-space spanned by $[x=\dot{u},y=u,z=u(t-\tau)]$ the dynamical evolution obeys the constrain $\dot{u}-f(u,u(t-\tau)) = 0$ and thus occurs in a two-dimensional manifold \cite{bunner1996recovery,bunner1997recovery,voss1997reconstruction,hegger1998identifying,bunner2000reconstruction,Bunner2000}. 

{To investigate the analogy between 1D SES and TDS, we analyze three systems: a 1D SES, a scalar TDS and a non-scalar TDS. 
The 1D SES is a reaction-diffusion bistable system. 
The scalar TDS is a bistable system with linear delayed feedback.
The non-scalar TDS is a model of two coupled lasers, with cross-delay terms.}
{We reconstruct their dynamics in the corresponding 3D pseudo-space: for the two TDS we use the same $[x, y, z]$ as in\cite{bunner1996recovery,bunner1997recovery}; for the 1D SES we define appropriated $[x, y, z]$, see Eq.~(\ref{eq:xyz_ses}).}

{We show that the evolution of these three systems in the 3D pseudo-space is well described by the equation $z = \mathcal{F}(x, y)$, where $\mathcal{F}$ is the same polynomial function.}
While this is expected for the 1D SES and for the scalar TDS (because of the way $x, y, z$ and $\mathcal{F}$ are defined), it is not expected for the coupled laser system (because of the complex structure of the model, which includes cross-delay-coupling terms among complex variables).
We end with a discussion of the limitations that in practice apply to the analogy between the evolution of 1D SES and TDS in the $[x, y, z]$ pseudo-space.

\section{Models}

\subsection{Spatially extended system (SES)}
\begin{figure}[tb]
  \centering
  \includegraphics[width=.389\textwidth]{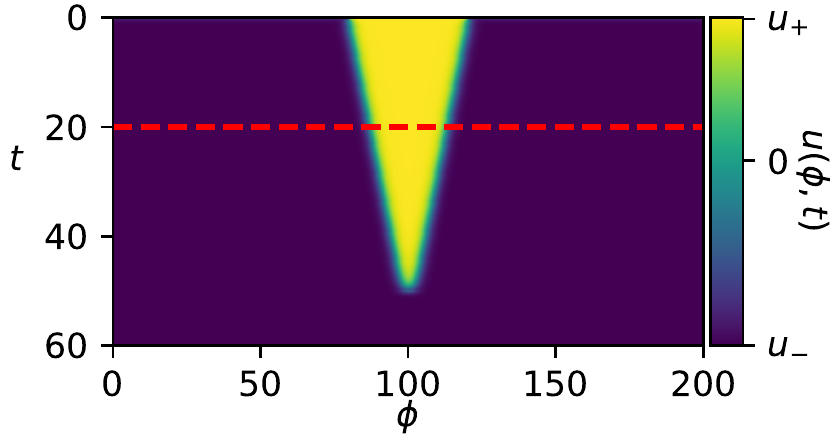}
  \caption{Space-time representation of the dynamics of the SES
model, Eqs. (\ref{eq:rds}) and (\ref{eq:F}) with parameters $\alpha=0.3$ and $D=4$.
{The dashed horizontal line indicates the time used in Fig.~\ref{fig:ses_phase}}.}
  \label{sds_num}
\end{figure}

The equation describing a reaction-diffusion 1D SES with a state variable $u(\phi,t)$ and potential $V(u)$ is:
\begin{equation}
    \partial_t u = F(u)  + D\partial_\phi^2 u, 
    \label{eq:rds}
\end{equation}
where $D$ is the diffusion coefficient and {$F (u)$ is the drift
force that we choose as}

\begin{eqnarray}
    F(u) &=& -dV(u)/du = -u(u +1+ \alpha)(u - 1)\nonumber\\
		     &=& -u^3 -\alpha u^2 + u(1+\alpha).
		\label{eq:F}
\end{eqnarray}
Here $V(u)$ is a double-well potential and $\alpha$ is the asymmetry of the potential. The model has steady states at $u=[0, u_+, u_-]$ with $u_+=1$ and $u_-=-\alpha - 1$. 

The model was simulated with spatial, $\Delta \phi$, and temporal, $\Delta t$, steps of $0.01$ and $0.0001$ respectively, and periodic boundary conditions.
A typical evolution of the system in time is presented in Fig.~\ref{sds_num}, where the color code represents the state variable $u(\phi,t)$ (the brightest color indicates $u_+$ and the darkest, $u_-$). Starting from an initial rectangular function with values $u_{+}$ and $u_{-}$, we see that the system evolves in time towards the lower state of the potential, $u_-$.

\begin{figure}[tb]
  \centering
    \includegraphics[width=0.4\textwidth]{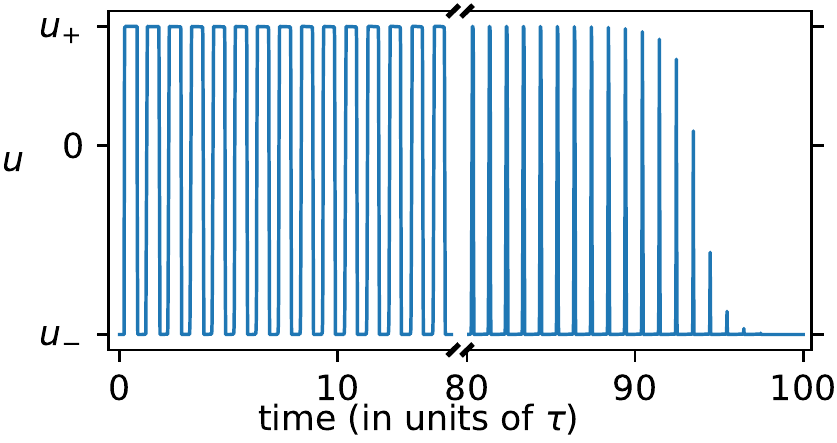}
  \caption{Time series of the TDS, Eq.~(\ref{eq:delay}), simulated with parameters $\alpha=2.5$, $\gamma=25$, $\tau=5$ and an integration step of $\Delta t=0.0001$. {The horizontal axis covers the intervals 0--10$\tau$ and 80$\tau$--100$\tau$ to make more clear the change in the oscillations. The temporal evolution in the pseudo-space, of the first pulse in the interval 80$\tau$--100$\tau$, is shown in Fig.~\ref{fig:fig_3dtds}.}}
  \label{ts_dst}
\end{figure}

\subsection{Time delayed system (TDS)} %.numerical delayed system (NDS).

\begin{figure}[tb]
  \centering
 \includegraphics[width=.4\textwidth]{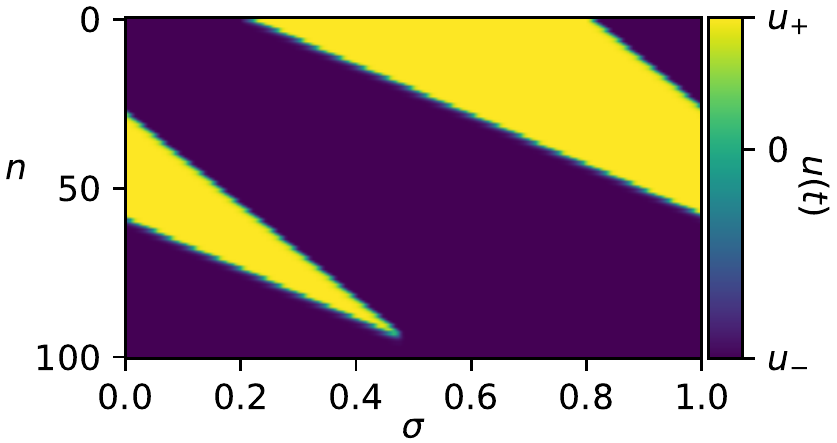}
  \caption{Space-time representation of the time series is shown in Fig.~\ref{ts_dst}: $u(t)$ with $t=n\tau + \sigma$ is plotted in color code vs. $n$ (the pseudo time) and $\sigma$ (the pseudo space).}
  \label{dst_num}
\end{figure}

We consider a scalar TDS that has the same potential as the 1D SES, and a linear feedback term with delay  $\tau$. Such system with a state variable $u(t)$ is described by:
\begin{equation}
  d{u}/dt = F(u) + \gamma u_{\tau}  %\\
  \label{eq:delay}
\end{equation}
where $F$ is given by Eq.~(\ref{eq:F}), $u_\tau = u(t-\tau)$, and $\gamma$ and $\tau$ are the strength and the delay of the feedback. This system has steady states at $(u_{0},u_{\pm})=\left [ 0, \quad (-\alpha \pm  \sqrt{(\alpha+2)^{2} + 4 \gamma })/2\right ]$. In order to integrate
Eq.~(\ref{eq:delay}) it is necessary to specify an initial function on the interval $[-\tau,0]$.
A typical solution obtained by using an initial rectangular function with values $u_{+}$ and $u_{-}$ is displayed in Fig. \ref{ts_dst}  {(here $\tau = 5$ and the integration step is $\Delta t = 0.0001$, which gives $50000$ steps bins in the interval $0-\tau$)}.
We can see that as time evolves, the time intervals during which the system remains in the higher state of the potential become gradually smaller until the system reaches the lower state of the potential. This behavior, characteristic of systems described by delay differential equations and referred to as metastability~\cite{nizette2003front,erneux2009applied}, is the equivalent to the propagation and annihilation of fronts in SES (leading eventually to a single phase). The corresponding space-time representation is displayed in Fig.~\ref{dst_num}. Here time is expressed as 
\begin{equation}
t=n\tau + \sigma
\label{eq:pseudo_time}
\end{equation} 
where $n$ is an integer number that plays the role of time, and $\sigma$ in $[0,\tau)$ plays the role of the space variable $\phi$ in the 1D SES.

\begin{figure}[tb]
  \centering
 \includegraphics[width=.4\textwidth]{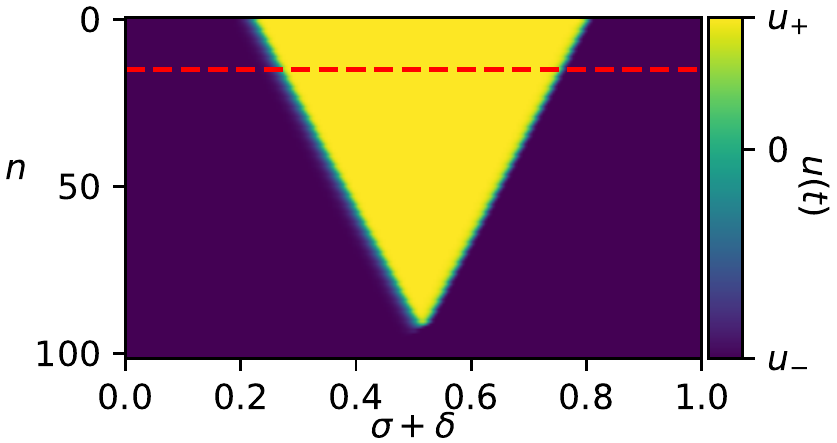}
  \caption{ Space-time representation of the TDS, after removing
the drift, { which is done by adjusting $\delta$ such that the pattern is vertically symmetric. Here $\delta = 26\Delta t$. The dashed horizontal line indicates the value of $n$ used in Fig.~\ref{fig:new_step4}(b).}}
  \label{dst_num_drif}
\end{figure}

The similarity between the space-time representation of the 1D SES and of the TDS is not obvious when comparing Figs.~\ref{sds_num} and \ref{dst_num}, because there is a linear drift in the spatio-temporal representation of the TSD, which is not present in the 1D SES. 
This drift, which is due to the fact that the TDS oscillation period is slightly larger than $\tau$, can be removed by defining $\sigma$ in $[0, \tau + \delta)$. 
{The value of $\delta$ (which is due to the system's finite response time to the feedback perturbation) can be estimated numerically: for the parameters in Fig. \ref{dst_num}, $\delta = 26\Delta t$.}
The resulting TDS space-time representation is shown in Fig.~\ref{dst_num_drif}, where now we note that, without the drift, there is a remarkable similarity with the space-time representation of the 1D SES, shown in Fig.~\ref{sds_num}.

\subsection{Coupled lasers system (CLS)}


As a more complicated time-delayed system, we consider two identical lasers, with symmetric, polarization-rotated optical coupling~\cite{masoller2011bifurcation,masoller2013two}. The model equations are
 
\begin{eqnarray}
 \frac{dE_{x,i}}{dt}  &=& k\left( 1+j\psi \right)\left( g_{x,i}-1 \right)E_{x,i}+\sqrt{\beta_{sp}}\xi_{x,i}, \\
 \frac{dE_{y,i}}{dt}  &=& j\Delta E_{y,i} + k\left( i + j\alpha_l \right)\left( g_{y,i} -1 -\beta \right)E_{y,i} \\
 &&+ \eta E_{x,3-i}\left( t - \tau \right)e^{-j\omega_{0}\tau} + \sqrt{\beta_{sp}}\xi_{y,i}, \nonumber\\ 
 \frac{dN_{i}}{dt}  &=&  \varepsilon_{N}\left( \mu -N_{i} - g_{x,i}I_{x,i}-g_{y,i}I_{y,i} \right)
\label{eq:DCSL}
\end{eqnarray}
where $i = [1,2]$ denote the two lasers, $E_{x,i}$ and $E_{y,i}$ are the orthogonal linearly polarized complex field amplitudes (the intensities being $I_{x,i}=|E_{x,i}|^2$ and $I_{y,i}=|E_{y,i}|^2$ respectively), and $N_i$ is the carrier density, of the i-th laser. $\omega_{0}$ is the emission frequency of the lasers; when the lasers are uncoupled the two frequencies are the same, and are equal to the frequency of the $x$ polarization that it is taken as the reference frequency. {$\Delta$} is the frequency detuning between the $x$ and $y$ polarizations. $g_{x,i}$ and $g_{y,i}$ are the gain coefficients that include self- and cross-saturation. Other parameters are: $k$ is the field decay rate, {$\varepsilon_N$} is the carrier decay rate, {$\alpha_l$} is the linewidth enhancement factor, $\beta$ is the linear loss anisotropy, $\beta_{sp}$ is the noise strength, $\xi_{x,i}$ and $\xi_{y,i}$ are uncorrelated Gaussian white noises and $\mu$ is the pump current parameter. 

The coupling strength is $\eta$ and the flight time between
the lasers (delay time) is $\tau=L/c$, with $L$ being the distance between the lasers and $c$ the speed of light. We note that polarization-rotated coupling means that cross-time-delayed terms are included in the rate equations of {two} complex variables: $E_{y,1}$ and $E_{y,2}$. 

\begin{figure}[tb]
  \centering
  \includegraphics[width=.3390\textwidth]{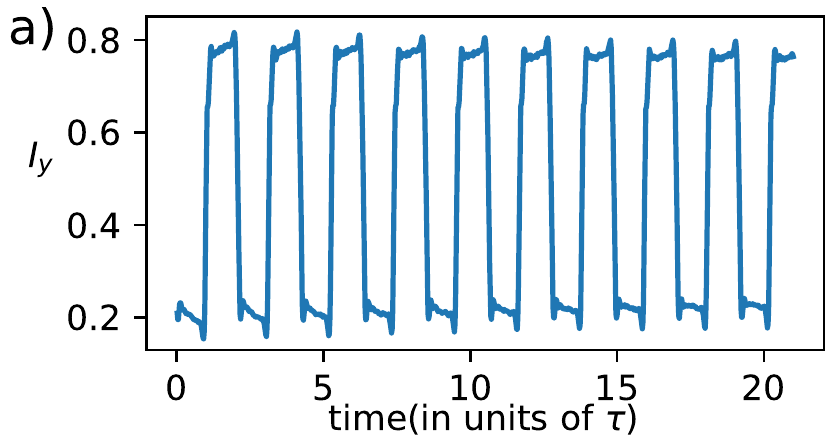}
  \includegraphics[width=.3390\textwidth]{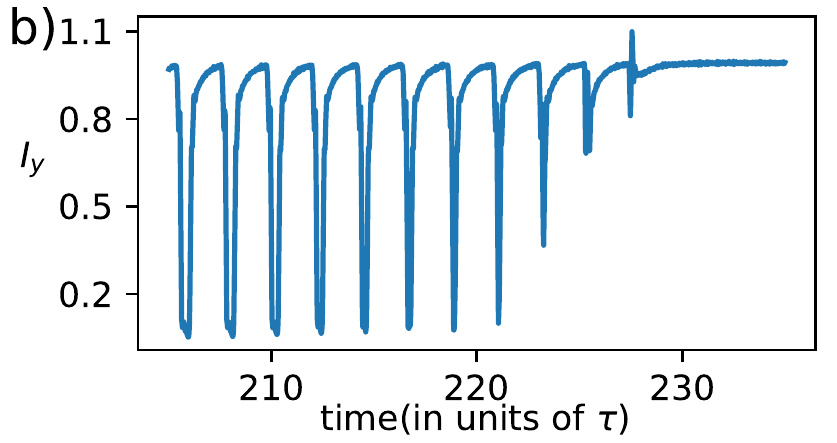}
  \caption{ Time series generated by simulation of the {CLS}
system. The intensity of the $y$ polarization of laser 1, $I_{y}=~|E_{y,1}|^{2}$, is plotted vs. time. The coupling strength is {(a)} $\eta = 36$ ns$^{-1}$ and {(b)} $\eta = 37$ ns$^{-1}$, other model parameters are: $k = 300$ ns$^{-1}$, $\mu = 2$, $\alpha_l = 3$, $\gamma_{N} = 0.5$ ns$^{-1}$, $\beta = 0.04$,
$\beta_{sp} = 10^{5}$ ns$^{-1}$ and $\tau = 3$ ns.}
  \label{ts_dcsl}
\end{figure}

Figure~\ref{ts_dcsl} displays  the temporal evolution of the intensity of the $y$ polarization of one of the lasers, where square-wave oscillations with period $2\tau$ are observed.
They are due to the polarization-rotated coupling and can either be stable or metastable, depending on the parameters\cite{masoller2011bifurcation,masoller2013two}.

\section{STATE SPACE RECONSTRUCTION}

It is possible to re-write the equation governing the
evolution of the 1D SES, Eq.~(\ref{eq:rds}) as
\begin{equation}
    z = \mathcal{F}(x,y) = F(x)  + Dy
		\label{eq:ses}
\end{equation}
where 
\begin{equation}
x=u,\quad y=\partial_\phi^2 u,\quad z=\partial_t u. 
\label{eq:xyz_ses}
\end{equation}
As $u$ is a function of the spatial variable $\phi$ and time, the pseudo coordinates ($x$, $y$, $z$) are also function of space and time. 

We can write the equation governing the evolution of
the TDS, Eq.~(\ref{eq:delay}), in the same form,
\begin{equation}
  z = \mathcal{F}(x,y) = F(x)  + \gamma y  %\\
		\label{eq:tds}
\end{equation}
where now
\begin{equation}
x = u,\quad y = u_\tau,\quad z=du/dt. 
\label{eq:xyz_tds}
\end{equation}
Using (\ref{eq:F}) $\mathcal{F}$ can be written as
\begin{eqnarray}
  \mathcal{F}(x,y) &=& A_0 +  A_1 x + A_2  x^2 + A_3 x^3 + A_4 y\nonumber \\
									 &+& A_5 xy + A_6 x^2 y.
\label{eq:manifold}
\end{eqnarray}
where 
\begin{eqnarray}
  A_{0} &=&  0,\nonumber \\
  A_{1} &=& 1+\alpha,\nonumber \\
  A_{2} &=& -\alpha,\nonumber \\
  A_{3} &=& -1,\nonumber \\
  A_{4} &=& D \mbox { for the SES,} \nonumber \\
  A_{4} &=& \gamma \mbox { for the TDS,} \nonumber \\
  A_{5} &=& A_6 = 0.
\label{estana_ses}
\end{eqnarray}

Therefore, the dynamics of both, the 1D SES and the TDS, are described in a 3D pseudo-space, by the equation $z = \mathcal{F}(x,y)$, where $\mathcal{F}$ is given in Eq.~(\ref{eq:manifold}). We hypothesize that the coupled laser system can also be described in a similar way.
Thus, {in order to test this hypothesis, we perform the following steps} to reconstruct the evolution {of each system in its own} pseudo-space:

\begin{enumerate}
\item First, we simulate the 1D SES and represent the dynamics in the pseudo-space using  $\left[x,y,z\right]=\left[u, \partial^2_\phi u , \partial_t u \right]$ as variables. 
Then, we fit the trajectory to $z = \mathcal{F}(x,y)$ and compare the fitted parameters with the theoretical ones, Eq.~(\ref{estana_ses}). 

\item  We repeat the procedure for the TDSs, using as variables $\left[x,y,z\right]=\left[u, u_\tau, du/dt \right]$. For the coupled laser system, the scalar variable used is the $y$-intensity of one laser, $u=I_{y}$.
\end{enumerate}

Numerically, all derivatives were estimated by the 2nd order finite difference. 
The parameters of the function $\mathcal{F}(x,y)$ were estimated by fitting $z = \mathcal{F}(x,y)$, with $x$, $y$ and $z$ computed as Eq.~(\ref{eq:xyz_ses}) for the 1D SES or as Eq.~(\ref{eq:xyz_tds}) for the TDSs. 
The \texttt{scipy.optimize.curve\_fit}\cite{scipy} algorithm was used to perform a non-linear least squares fit of the function $\mathcal{F}$, Eq.~(\ref{eq:manifold}), to the data points.

\begin{figure}[tb]
  \centering
  \includegraphics[width=.48\textwidth]{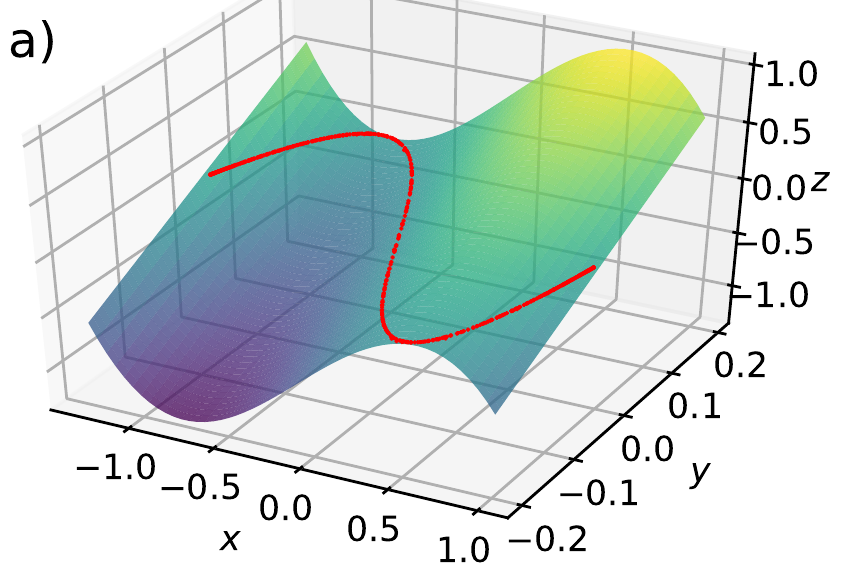}
  \includegraphics[width=.39\textwidth]{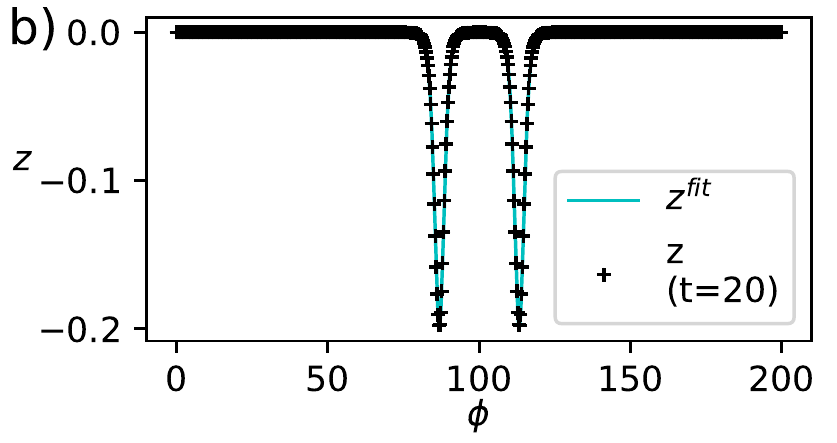}
  \caption{
  (a) Representation of the SES dynamics (shown in Fig.~\ref{sds_num}) in the pseudo phase space ($x=u$, $y=\partial_{\phi}^{2}u$, $z=\partial_{t}u$). 
  The surface in color indicates the value of $\mathcal{F}(x,y)$ with the parameters given in Table~\ref{tab:tab1}, {the red points represent the trajectory computed from the time series $u(t)$. A 3D representation of the attractor can be found in \cite{figgif}}.  (b) Comparison between the numerical time series {(along the dashed horizontal line in Fig.~\ref{sds_num}) and the fitted time series}.}
  \label{fig:ses_phase}
\end{figure}

\section{Results}

Figure~\ref{fig:ses_phase} displays the evolution of the 1D SES~(that~was shown in Fig.~\ref{sds_num}), now represented in the pseudo-space $[x,y,z]$. 
The fit of the trajectory to $z = \mathcal{F}(x,y)$ gives the parameters listed in Table~\ref{tab:tab1}. 
These values are in excellent agreement with the theoretical values given by Eq.~(\ref{estana_ses}), which, for $\alpha = 0.3$ and $D = 3$ are also listed in Table~\ref{tab:tab1}. 
The small error (less than $3\%$) is attributed to the numerical estimation of the spatial and temporal derivatives.
{In Fig.~\ref{fig:ses_phase}(b) the two negative spikes can be understood as follows: in Fig.~\ref{sds_num} at time $t$=55, $z=\partial u/\partial t=0$ except at the two boundaries of the yellow region, where $z < 0$ because in both boundaries $u$ decreases (note that time increases downwards and z measures the ``vertical'' variation of $u$, from $u_{+}$ to $u_{-}$).}

\begin{figure}[tb]
  \centering
  \includegraphics[width=.480\textwidth]{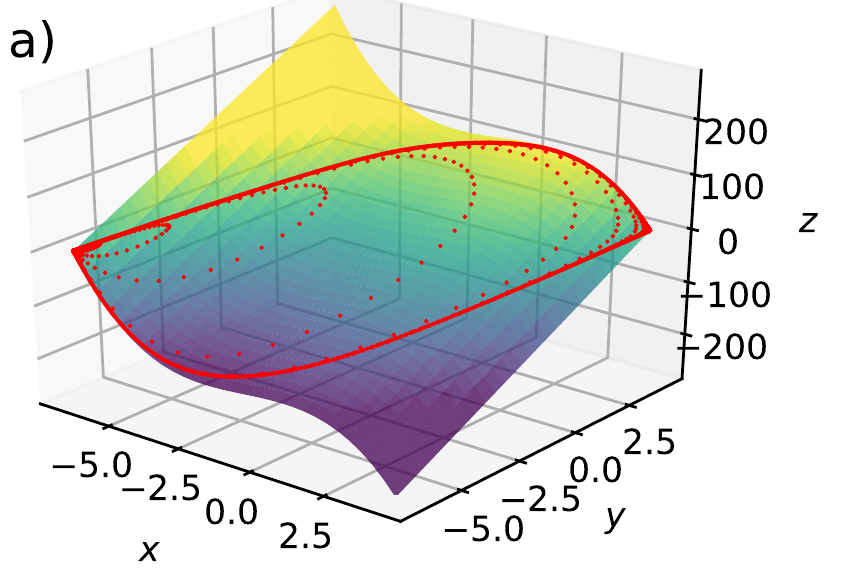}
  \includegraphics[width=.410\textwidth]{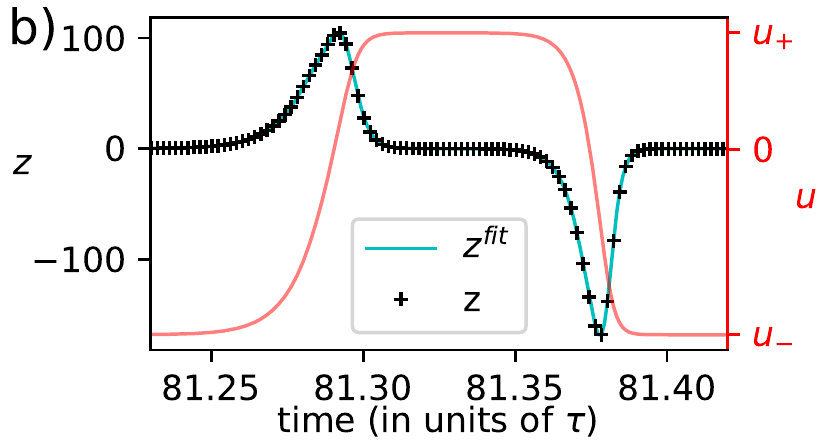}
  \caption{(a) Representation of the TDS in the phase space ($x=u$, $y=u_\tau$, $z=du/dt$). The surface in color indicates the value of $\mathcal{F}(x,y)$ with the parameters given in Table~\ref{tab:tab1}, {the points represent the trajectory computed from the time series $u(t)$. A 3D representation of the attractor can be found in \cite{figgif}. (b) Comparison between the numerical and the fitted time series. Here the right vertical axis displays $u(t)$, the time interval corresponds to the first pulse in the interval 80$\tau$--100$\tau$ shown in Fig.~\ref{ts_dst}}.}
  \label{fig:fig_3dtds}	
\end{figure}

Figure~\ref{fig:fig_3dtds} displays the evolution of the TDS (that was shown in Fig.~\ref{ts_dst}), now represented in the pseudo-space $[x,y,z]$.
The fit of the trajectory to $z = \mathcal{F}(x,y)$ gives the parameters listed in Table~\ref{tab:tab1}. 
We again obtain an excellent agreement with the theoretical values that correspond to $\alpha = 2.5$ and $\gamma = 25$ (the error is less than $0.5\%$). 

\begin{figure}[ht]
  \centering
  \includegraphics[width=.480\textwidth]{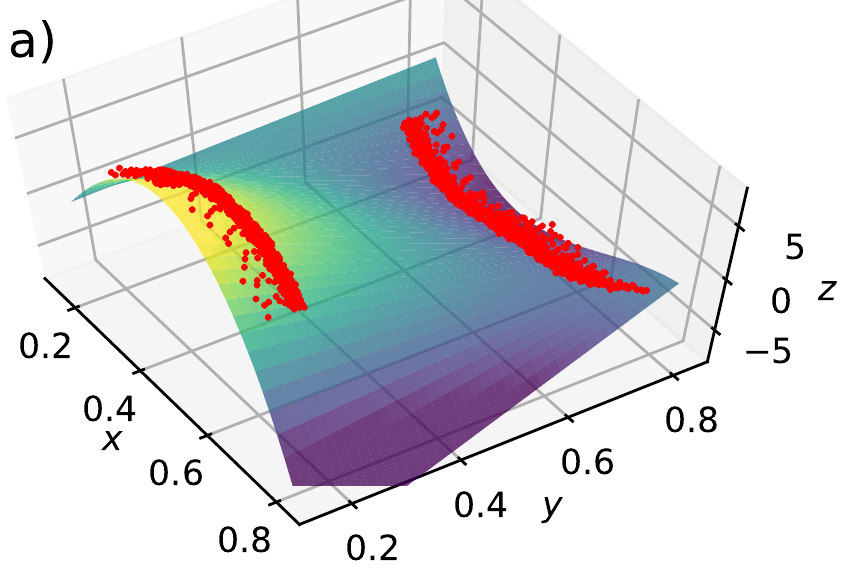}
  \includegraphics[width=.480\textwidth]{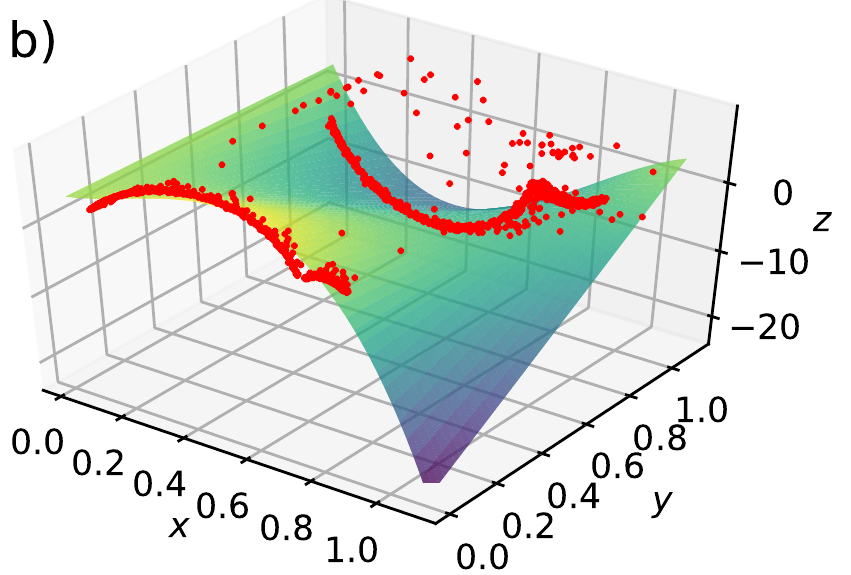}
  \includegraphics[width=.390\textwidth]{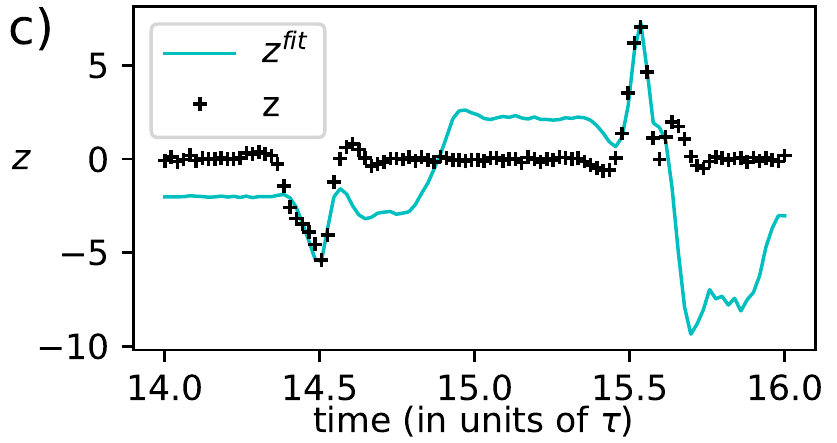}
  \includegraphics[width=.390\textwidth]{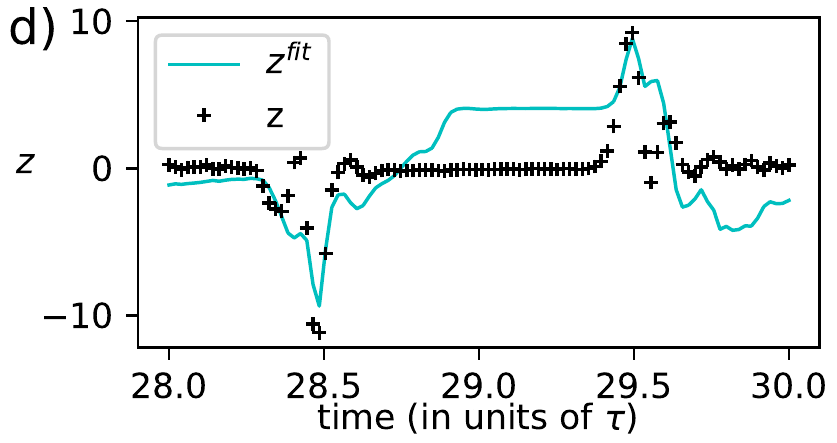}
  \caption{(a) Representation of the evolution of the coupled laser system (CLS, the time series shown in Fig.\ref{ts_dcsl}) in the phase space $[x=u,y=u_{\tau},z=du/dt]$ with $u=I_{y}$, together with the function $\mathcal{F}(x,y)$. A 3D representation of the attractor can be found in \cite{figgif}. (b) Comparison between the numerical and the fitted time series.}
  \label{fig:4.1}
\end{figure}

{The results presented so far, expected due to the way
$x,y,z$ and $\mathcal{F}$ have been defined, have allowed us test the
accuracy of the fitting program.}
For the coupled laser system, because the model structure is more complicated, we did not expected to obtain a very good fit with the same polynomial function that fits  the scalar TDS.
Nevertheless, for the time series shown in Fig.~\ref{ts_dcsl}, as shown in Fig.~\ref{fig:4.1}, we can fit the trajectory to $z = \mathcal{F}(x,y)$ with the parameters given in Table~\ref{tab:tab1}. 
In this case no relation could be inferred between the fitted parameters and the model's parameters. 
{A possible way to improve the fitting is to add higher order terms to the function $\mathcal{F}(x,y)$, or to test other functional relationships.}

To further test the analogy between the 1D SES and the TDS, we addressed the following question: can we use the $[x,y,z]$ definitions for the 1D SES given in Eq.~(\ref{eq:xyz_ses}) to reconstruct the state space of the TDS?

\begin{figure}[ht]
   \centering
   \includegraphics[width=.480\textwidth]{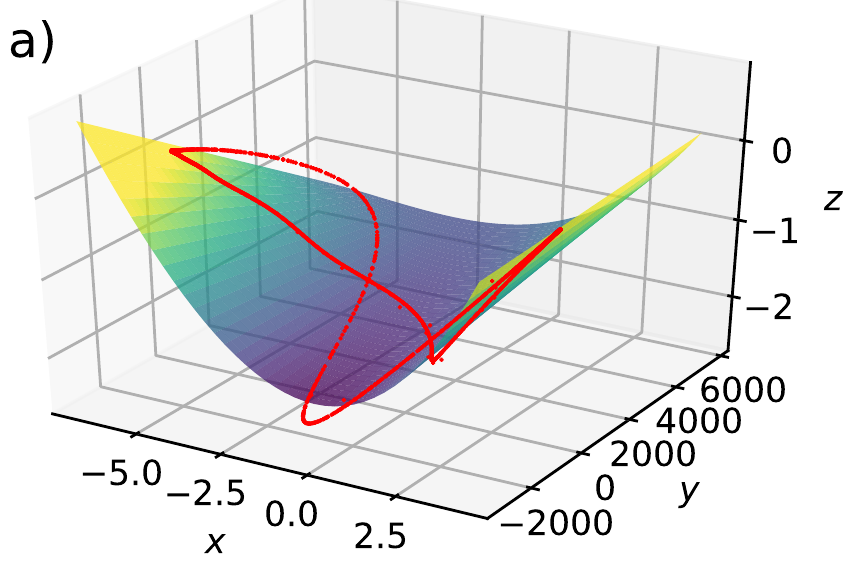}
   \includegraphics[width=.390\textwidth]{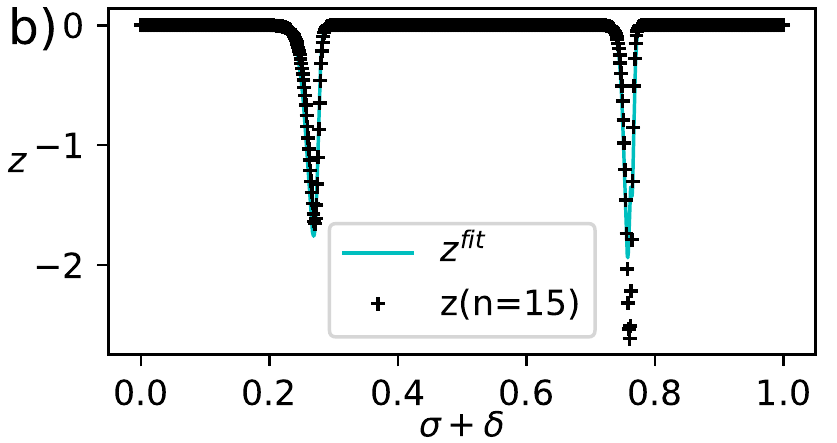}
   \includegraphics[width=.390\textwidth]{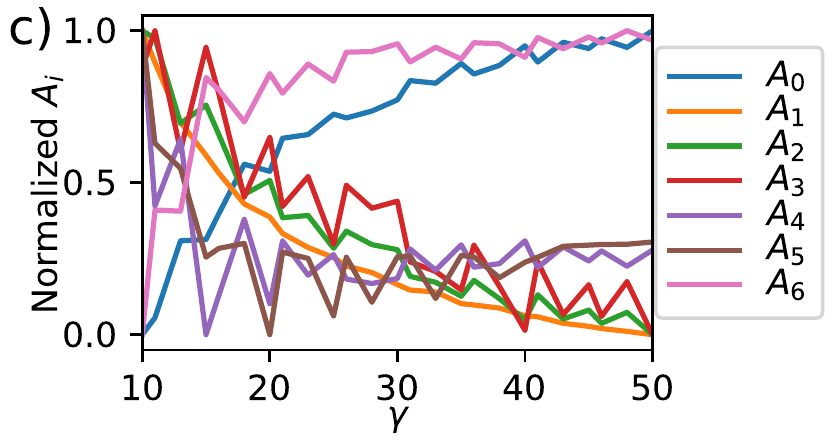}
   \caption{(a) Representation of the dynamics of the TDS (the time series shown in Fig.~\ref{ts_dst}), in the phase space $[x,y,z]$, with $x$, $y$, and $z$ defined for the 1D SES, Eq.~(\ref{eq:xyz_ses}). A 3D representation of the attractor can be found in \cite{figgif}. (b) Comparison between the numerical time series {(along the dashed horizontal line in Fig.~\ref{dst_num_drif}) and the fitted time series.} (c) Normalized fitted parameters ($A_i$) for different values of $\gamma$. }
   \label{fig:new_step4}
 \end{figure}

To address compute $[x,y,z]$ according to Eq.~(\ref{eq:xyz_ses}) it is necessary to calculate the partial derivatives, replacing $t$ with the ``time variable'', $n$, and $\phi$ with the ``space variable'', $\sigma+\delta$, in the space-time representation of the TDS, after removing the drift (shown in Fig.~\ref{dst_num_drif}). 
Numerically, the calculation of the partial time derivative can introduce errors because, by definition, $n$ is an integer number and thus, the smallest time step for calculating the derivative is $\Delta n$ = 1. 
Nevertheless, as shown in Figs.~\ref{fig:new_step4}(a) and (b), it is possible to obtain a reasonably good fit of the trajectory, with the fitted parameters also listed in Table~\ref{tab:tab1}.
To investigate if there is a relation between the fitted parameters and the $\gamma$ parameter of the TDS, in Fig.~\ref{fig:new_step4}(c) we show how the values of the fitted parameters change with $\gamma$. 
For easy comparison ({because the $A_{0}-A_{6}$ parameters can have very different values as seen in Table~\ref{tab:tab1}}), the fitted values are normalized with respect to the minimum and maximum values. 
We observe that all parameters vary with $\gamma$: $A_{0}$ and $A_{6}$ increase while the others decrease. {While $A_0$ and $A_6$ are exactly equal zero for the SES and for the scalar TDS, they can be non-zero for other systems, and as shown in Table 1, they are not zero for the CLS and also, when the TDS is fit as a SES. In this case (TDS2 in Table~\ref{tab:tab1}) some parameters have very small values and to analyze the significance of the fitted values, we divided the time series in windows of different lengths and fitted the parameters in each window. The relative error of the fitted values found was very small (about 1-2\% depending on the window length). To analyze the significance of each individual parameter $A_i$, we assumed $A_i=0$, fitted all the other $A_j$s, and calculated the relative change in $A_j$ when $A_i\ne0$. We found that setting $A_4$ or $A_5=0$ or $A_6=0$ produced small relative variations in the values of $A_0-A_3$ (less than 10\%) while setting $A_0$ or $A_1$ or $A_2$ or $A_3$ equal to zero resulted in large variations of the other parameters (larger than 100\%). While a more detailed study is need to find the combination of parameters that gives the best fit for the TDS2, the analysis suggests that a good fit can also be obtained by setting either $A_4$, $A_5$ or $A_6$ equal to zero.}

A conceptual limit of the SES-TDS analogy in the pseudo phase space comes from temporal causality: the variable $y=u(t-\tau)$ of the TDSs is a function of time, and therefore, its evolution is constrained by temporal causality; in contrast, for the 1D SES, $y$ is the second order spatial derivative, $y=\partial_{\phi}^{2} u$ and therefore, its evolution is not restricted by causality.

\newcommand{\mc}[3]{\multicolumn{#1}{#2}{#3}}

\begin{table*}[tb]
 \footnotesize
\caption{Theoretical and fitted parameters obtained for the 1D SES, the TDS, and the coupled laser system. CLS: $\eta=36$~ns$^{-1}$,
CLS2: $\eta=37$~ns$^{-1}$, TDS2: the TDS fitted as a SES. $-$ indicates that the theoretical values are unknown.
}
\label{tab:tab1}
\begin{tabular}[c]{|c|c|c|c|c|c|c|c|c|c|c|c|c|c|c|}
\hline
% \multirow{ 2}{*}{Systems} & \mc{14}{c|}{Theoretical $|$ Fitted}\\
 {System} & \mc{14}{c|}{Theoretical $|$ Fitted}\\
\cline{2-15}
% × & \mc{2}{c|}{} &  &  &  &  & & \\
&\mc{2}{c|}{$A_0$}   & \mc{2}{c|}{$A_1$} & \mc{2}{c|}{$A_2$} &  \mc{2}{c|}{$A_3$} & \mc{2}{c|}{$A_4$} & \mc{2}{c|}{ $A_5$} & \mc{2}{c|}{$A_6$} \\\hline
SES  & 0 & -1$\times10^{-6}$& 1.3 & 1.29	& -0.3 & -0.29 	& -1 & -0.99 	& 3 & 2.99 	& 0 & 7$\times10^{-7}$	& 0 & 1$\times10^{-7}$ \\
TDS  & 0 & -3.3$\times10^{-2}$	& 3.5 & 3.5 	& -2.5 & -2.5 	& -1 & -1 	& 25 & 25 	& 0 & 9$\times10^{-3}$ 	& 0 & 3$\times10^{-3}$ \\
CLS  & $-$ & -13.7	& $-$ & 92 	& $-$ & -19 	& $-$ & -124 	& $-$ & 36.1 	& $-$ & -273 	& $-$ & 312 \\
CLS2 & $-$ & 3.8	& $-$ & 8 	& $-$ & 36 	& $-$ & -61 	& $-$ & 10 	& $-$ & -93 	& $-$ & 105 \\
TDS2 & $-$ & -1.761	& $-$ & 8.1$\times10^{-2}$ 	& $-$ & 6.96$\times10^{-2}$ 	& $-$ & 2.5$\times10^{-3}$ 	& $-$ & 7$\times10^{-5}$	& $-$ & 1$\times10^{-5}$	& $-$ & -5$\times10^{-6}$\\
\hline
\end{tabular}
\end{table*}

\section{Conclusions}

We have studied the analogy between a time-delayed system (TDS) and a one-dimensional spatially extended system (1D SES) by considering the particular examples of a 1D reaction-diffusion SES, and a bistable scalar system with a linear feedback term (TDS).
We have also considered a {non-scalar TDS: a model of two symmetrically coupled lasers, with cross-delay terms in two} complex variables.

{We have shown that the evolution of these systems can be described, in a three dimensional pseudo-phase space, by $z = \mathcal{F}(x,y)$, where $\mathcal{F}(x,y)$  is the same polynomial function, Eq.~(\ref{eq:manifold}), and the variables $x$, $y$ and $z$ are defined as in Eq.~(\ref{eq:xyz_ses}) for the 1D SES and Eq.~(\ref{eq:xyz_tds}) for the scalar TDS and non-scalar TDS.
For the 1D SES and for the scalar TDS the values of the fitted parameters of $\mathcal{F}(x,y)$ were in excellent with the theoretical values (as expected due to the way $x, y, z$ and $\mathcal{F}$ were defined); for the non-scalar TDS (i.e., the coupled lasers system) a reasonably good fit of the function $\mathcal{F}$ was obtained, but} no relation was found between the fitted parameters and the model parameters.

While this approach could in principle be applied to any TDS or SES, a main limitation for a successful reconstruction of the phase space is the estimation of the parameters of the function $\mathcal{F}$, which, in systems with more complicated governing equations will likely not be limited to a low-order polynomial. Therefore, a main challenge is a reliable estimation of the parameters of $\mathcal{F}$, when many parameters need to be estimated.

\section*{Acknowledgments}
This work was supported in part by ITN NETT (FP7 289146), the Spanish MINECO (FIS2015-66503-C3-2-P) and the program ICREA ACADEMIA of Generalitat de Catalunya. 
The authors acknowledge G. Giacomelli and F. Marino the introduction to the problem that motivated this work and very useful discussions. 
C.Q. also acknowledges G. Giacomelli and F. Marino for their hospitality during his visit to the Istituto dei Sistemi Complessi, Florence, where this work started.

% \hspace{50cm}

\bibliography{STAnalogy}

\end{document}